# Keeping light pollution at bay: a red-lines, target values, top-down approach

**Salvador Bará[1], Fabio Falchi[1,2,*], Raul C. Lima[3,4], Martin Pawley[5]**

[1]*Dept. de Física Aplicada, Universidade de Santiago de Compostela, 15782 Santiago de Compostela, Galicia*

[2]*Istituto di Scienza e Tecnologia dell'Inquinamento Luminoso (ISTIL), 36016 Thiene, Italy.*

[3]*Escola Superior de Saúde do Politécnico do Porto, 4200-072 Porto, Portugal*

[4]*Centro de Investigação da Terra e do Espaço da Universidade de Coimbra, Almas de Freire – Sta. Clara 3040-004 Coimbra, Portugal*

[5]*Agrupación Astronómica Coruñesa Ío, 15005 A Coruña, Galicia*

*Corresponding Author: F. Falchi (falchi@istil.it)

**Abstract**

The prevailing regulatory framework for light pollution control is based on establishing conditions on individual light sources or single installations (regarding features like ULOR, spectrum, illuminance levels, glare, ...), in the hope that an ensemble of individually correct lighting installations will be effective to somehow solve this problem. This "local sources" approach is indeed necessary, and shall no doubt be enforced; however, it seems to be clearly insufficient for curbing the actual process of degradation of the night, and for effectively attaining the necessary remediation goals. In this paper we describe a complementary (not substitutive) 'red-lines' strategy that should in our opinion be adopted as early as possible in the policies for light pollution control. This top-down approach seeks to set definite limits on the allowable degradation of the night, providing the methodological tools required for making science-informed public policy decisions and for managing the transition processes. Light pollution abatement should routinely be included as an integral part of any territorial management plan. A practical application case-study is described to illustrate these concepts.

# 1. Introduction

Artificial light is arguably one of the key inventions of humankind, and no doubt it has brought innumerable benefits to our society. However, as it generally happens with successful technical developments, it has also negative side-effects whose importance and extent are being revealed by a growing body of research. These detrimental effects are not limited to a specific field, but transversally affect the environment, the nocturnal landscape, the starry sky as a scientific and cultural global commons, and arguably, public health, among others (Aubé et al, 2020; Alamús et al, 2017; AMA 2012, 2016; Bará, 2016; Bennie at al, 2015; Bonmati-Carrion et al, 2014; Cho et al, 2015; Cinzano et al, 2001; Czeisler, 2013; Davies et al, 2015, 2015, 2016, 2020; Dobler et al, 2015, 2016; Falchi et al, 2016; Garcia-Saenz et al, 2018;  Gaston et al, 2013, 2014; van Grunsven et al, 2020; Haim and Portnov, 2013; Hölker et al, 2010, 2010b; Kyba et al, 2017; Longcore and Rich, 2004; Marin and Jafari, 2008; Rich and Longcore, 2006; Rybnikova and Portnov, 2016, 2018; Sanders et al, 2020). Being artificial light at night (ALAN) a pollutant, addressing its negative effects is becoming part of the public agenda in many countries of the world (Bará et al, 2019; Falchi et al, 2011; Falchi and Bará, 2020; Longcore et al, 2015; Schroer et al, 2020; Zielińska-Dąbkowska et al. 2020).

The current regulatory framework, where present, is significantly bent towards imposing controls upon individual light sources and installations, without sufficient regard to the accumulated detrimental effects they produce on the environmental parameters we effectively want to protect. In this paper we argue that these local control measures are necessary and shall be enforced, but they should be accompanied by a complementary top-down approach oriented to warrant that maximum admissible limits of deterioration are not surpassed and, if they were, definite remediation steps will be taken to revert to an acceptable situation.

We describe here some aspects of this approach, outlining the procedures for evaluating limit compliance and discuss how to circumvent some practical difficulties. According to the outcome of that evaluation, mid-term territorial management plans should be elaborated oriented either to ensure preservation, in case of compliance, or to achieve remediation, in case of failure. We will illustrate our comments with a case-study based on the Galician Atlantic Islands Maritime-Terrestrial National Park, which is a Starlight certified destination for astro-tourism (Starlight Foundation, 2015), in Galicia (Spain, European Union). The content of this example should not be taken literally; it is included here only to help the discourse. Many details and side-discussions will necessarily be omitted, leaving to the insightful wisdom of the readers the task of filling in the gaps. Needless to say, we provide these considerations as a perspective fully open to criticism, expecting that they might be helpful for fostering the necessary debate on how to address the challenges faced nowadays by all people and institutions concerned with the preservation of the night.

## 2. Local sources control: necessary, but not sufficient

The prevailing approach to light pollution mitigation is mainly based on the adoption of engineering and administrative measures upon the individual light sources or individual installations at street or district level. This approach seeks to reduce their excessive light emissions, especially the most detrimental ones, and is formulated basically (although not exclusively) in terms of intensive quantities, that is, indicators of relative performance. Some examples of these intensive quantities are the fraction of light emitted by a luminaire toward the upper hemisphere, the average number of lumen per square meter (lx) on the lit surfaces, the fraction of the light emitted by a lamp that effectively reaches the area intended to be illuminated (utilization factor), the luminous efficacy of the lighting system, measured in lm per W, or the relative spectral content of the light, usually described by CCT for visual purposes and by other metrics for non-visual applications (Bará et al, 2019b; Galadí-Enríquez, 2018; Lucas et al, 2014; Rea and Figueiro, 2016; Rea et al, 2012; Sánchez de Miguel et al, 2019).

This traditional approach has arguably allowed to reduce the rate of increase of light pollution, in comparison with a non-intervention scenario, and there is little doubt that it is a convenient and necessary component of any sensible environmental strategy. However, it also seems apparent that it falls short of providing the required tools to effectively control the negative effects of ALAN. There are two basic reasons for this insufficiency: on the one hand, tightening the relative emission limits (e.g. lumen per luminaire, lx on the streets, etc) does not imply by itself containing the overall light emissions (e.g. total emitted lm, total spectral radiance in $W/m^2/sr/nm$), which are the determining factor of the real extent of the environmental damage, and which currently increase at a sustained rate (Kyba et al, 2017); on the other hand, the relative limits established in most outdoor lighting recommendations, standards, and legal regulations, certainly were not designed for, nor sought to be consistent with, environmental, landscape, and public health preservation goals. They are supposed to satisfy the visual performance needs for vehicle drivers and pedestrians, but arguably they lack a sound and clear scientific rationale (for a review, see Fotios and Gibbons, 2018), they are strongly driven by the lighting industry needs, and most of them are outdated, proposing exaggerated values for the intended effect.

Whereas it is expected that the local sources approach could contribute to reduce the absolute deterioration of the environment, the link between its proposed measures and the actual reduction of the overall light pollution negative effects is not logically established nor very often verified.

## 3. A complementary strategy: setting immission limits

We propose the adoption of a complementary, non-substitutive approach to build a comprehensive light pollution control strategy. This consists of a classical immissions control approach, commonly incorporated in environmental management and public health regulations, based on (i) the specification of quantitative limits for the maximum allowable deterioration of the night environment, and (ii) the implementation of a top-down procedure to deduce the constraints that must be satisfied by the ensemble of intervening lighting installations to comply with these limits.

This complementary approach should provide the information required to effectively limit the overall accumulated emissions and should help reframing in a wider context the actual relative limits recommendations (e.g. % blue content of the lamps, etc), without substantially modifying them. The maximum acceptable limits included in this perspective are formulated in terms of absolute (extensive) rather than relative (intensive) physical quantities, and are usually specified at the location where the damage is produced (e.g. the maximum horizontal irradiance in a relevant ecological photometric band, eventually averaged within a protected area) rather than at the light source location (lx on the street).

As explained below, adopting maximum quantitative limits on the detrimental effects of ALAN logically requires the existence of an absolute limit to the overall light emissions from the surrounding territories (independently of whether it is made explicit or not in the regulations), such that, if this limit is exceeded, the negative effects will be larger than those we are willing to accept. This naturally leads to the need of adopting tight light emission caps, whose territorial allocation is a science-informed but essentially social and political issue. The logical link between limiting the effects and limiting the overall emissions is a necessary consequence of the physics of the propagation of light in the atmosphere (Aubé, 2015; Aubé et al, 2020; Bará and Lima, 2018; Bará et al, 2019c; Cinzano and Falchi, 2012; Falchi and Bará, 2020; Garstang, 1986, 1989, 1991; Kocifaj, 2016).

## 4. Basic elements of a red-lines, target values, top-down approach

By 'red-lines' we mean the explicit and quantitative limits that concerned stakeholders are not willing to exceed for the degradation of the night environment, that urge the need of adopting preventive measures to avoid reaching them, and of short-term remediation actions in case they are exceeded. In the latter case, the red-lines become the 'target values' of the remediation action. Both red-lines and target values refer to the same quantitative values; the difference in using these terms stems from whether the actual deterioration levels have already surpassed the limit of acceptability.

A red-lines or target values based top-down approach should include, among others, the following steps:

1. Defining the detrimental consequences to be addressed

2. Choosing the appropriate indicators and setting the red-line values

3. Evaluating compliance

4. Evaluating preventive or remediation options, according to the outcome of 3.

5. Allocating emission/reduction quotas and duties among intervening actors

6. Steering the transition process

We briefly discuss these steps in the paragraphs below. We illustrate them with some interspersed comments describing an actual but relatively simple example: the preservation of the night sky brightness in the Galician Atlantic Islands Maritime-Terrestrial National Park.

*4.1. Defining the detrimental consequences to be addressed*

A basic starting point of any strategy for light pollution control should be a clear specification of the detrimental effects to be addressed, and of the red-lines to be avoided (or the remediation target values to be achieved) at the places where the deterioration happens. Admittedly, this is not a trivial task, since we still lack a complete picture of the complex interrelations between ALAN and the natural world. This issue could be addressable in a foreseeable future in what regards human health (linear and non-linear interactions of artificial light with the complex biological system of a single species), and it is arguably more difficult to solve for the natural environment at large (linear and non-linear interactions of light intra- and inter-species in complex ecological networks). This lack of complete knowledge, however, is common to many polluting agents that we are controlling and abating, and should not prevent us from adopting some provisory limits. In the wise words of Sir Austin Bradford-Hill, key player in the recognition of tobacco smoke as a serious health hazard, in its classic 1965 discourse "The Environment and Disease: Association or Causation?" (Bradford-Hill, 1965)

> *"(…) All scientific work is incomplete - whether it be observational or experimental. All scientific work is liable to be upset or modified by advancing knowledge. That does not confer upon us a freedom to ignore the knowledge we already have, or to postpone the action that it appears to demand at a given time."*

There is however a particular field where the effects to be addressed can be confidently defined with little or no ambiguity: the artificial brightness of the night sky. Some first-class optical observatories have already undergone major losses on the night sky over them, others are currently losing it and others may still find jeopardized their ability to carry out the scientific observations for which they were built and equipped if the artificial brightness surpasses some definite instrument-dependent limits (Walker, 1970; Falchi and Bará, 2020). The effect to keep under control is then clear: the artificial radiance entering the photometric bands of their observing instruments, including, when appropriate, the naked human eye. The same can be applied to any dark site whose night skies are to be protected. As a case study let us resort here to one Tourism Destination in Galicia certified by the Starlight Foundation (Starlight Foundation, 2015) whose basic description is provided below.

> **Case study (1/6):** The Galician Atlantic Islands Maritime-Terrestrial National Park is an ensemble of four archipelagos located in the Galician west coast, a few miles off a densely populated and highly illuminated shoreline (Fig 1). Only the Ons island is permanently inhabited by persons other than the National Park staff, but the Park receives more than 400,000 visitors during the year, especially in the summer season, including numerous research groups developing field campaigns to study the rich ecosystems and endemic species of the islands. Since 2016 it is a certified Starlight Tourism Destination and there is an intense activity of stargazing carried out by the Park itself, support groups from the Galician universities, amateur astronomer associations, and an increasing number of recreational private firms. The prevailing atmospheric conditions of this location (low scattering, due generally to low aerosol content) allow for reasonably dark skies even if located close to highly pollutant population nuclei. However, if light pollution is not effectively controlled, these skies would be so compromised as to breach in a relatively short term the requirements to maintain the certification, if as of today they are not already jeopardized.

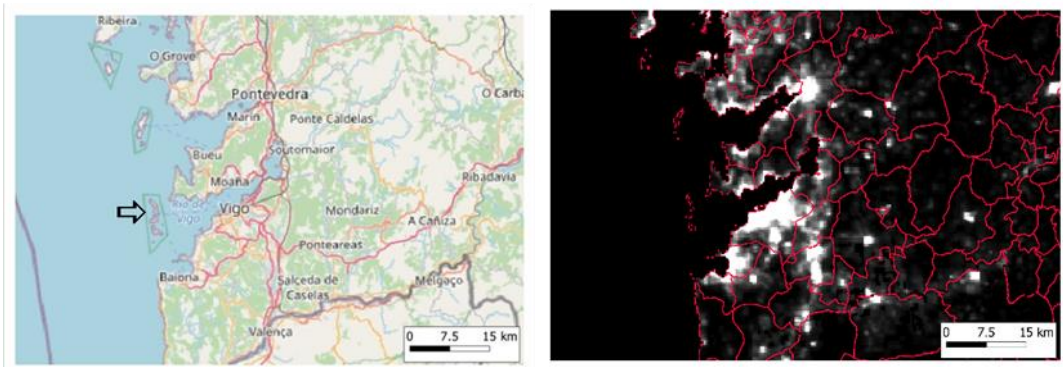

**Fig. 1.** Left: The Galician Atlantic Islands Maritime-Terrestrial National Park. The black arrow indicates the location of the Cíes Islands, whose night sky brightness is discussed in this case study. Right: VIIRS-DNB 2015 stable lights composite (Earth Observation Group, 2018), with superimposed borders of municipalities.

*4.2. Choosing the indicators and setting the red-lines*

The next step is choosing adequate quantitative indicators to measure the detrimental light pollution effects. As commented above, there is a pressing need of developing appropriate indicators for environment and public health. Some interesting steps have been done in what regards acute melatonin suppression, since a few recent models are able to predict with reasonable accuracy the percent control-adjusted reduction in circulating melatonin (Rea et al, 2005, 2012; Rea and Figueiro, 2016). Note that other quantities like the absolute exposure in the melanopic band (Lucas et al, 2014; CIE, 2018), or the relative MSI (Aubé et al, 2013) and G-index (Galadí-Enríquez, 2018) indicators would not be appropriate for this particular task, since they cannot provide by themselves a quantitative prediction of the true extent of the melatonin suppression under general exposure conditions. Relative environmental indicators are also available (Donners et al, 2018, Longcore et al, 2015, 2018), being a sound and sensible starting point that can be provisionally adopted while the required absolute ones are developed.

The situation is again somewhat easier to handle when it comes to artificial night sky brightness in observatories or dark sky destinations. The artificial sky radiance is routinely measured in observatories in multiple instrumental bands, see e.g Bessel (2005), and Casagrande and Vandenberg (2014). Although the red-lines to be adopted are not often explicitly stated, determining them should not present any essential difficulty: any seasoned astronomer can estimate the maximum amount of scattered moonlight that would render useless the observations within their instrument passband. As a matter of fact, the International Astronomical Union, in a seminal 1976 resolution, noted "with alarm the increasing levels of interference with astronomical observation resulting from artificial illumination of the night sky" urgently requesting "that the responsible civil authorities take action to preserve existing and planned observatories from such interference", offering to provide "information on acceptable levels of interference and possible means of control." (IAU, 1976, p.7). The IAU further recommended that the acceptable levels of interference by artificial illumination were limited "to a small fraction of the natural sky brightness" (ibid, p.30). In 1979, this "small fraction" was explicitly quantified, the IAU recommending that

> *"The increase in sky brightness at 45º elevation due to artificial light scattered from clear sky should not exceed 10 per cent of the lowest natural level in any part of the spectrum between wavelengths 300 and 1000 nm except for the spectral line emission from low pressure sodium lamps as set out in Recommendation 2 (...)"* (Cayrel, 1979, p 220).

The final value adopted for a red-line is always the result of a balanced assessment of different intervening factors, and in many cases other values could have been chosen as well. As pointed out by Cayrel, "The choice of a 10% contribution of artificial lighting to the natural background is, of course, somewhat arbitrary and is intended to mean that the background should not be significantly increased" (Cayrel, 1979).

As a general comment, the levels of accuracy and precision required for scientific work and for drafting science-informed public policies are usually widely different. This could create a certain degree of discomfort, since we are used in our everyday work in science to getting right the n-th decimal place in our results with very low uncertainty, as a necessary condition for ascertaining the validity of some theoretical models, or for adequately monitoring the long-time trends of the sky brightness in any photometric band. Besides, the red-lines are commonly established for a given set of conditions (e.g. standard human observers, typical aerosol loads, ...), and very likely may fall short of capturing all the variability inherent to human and natural affairs. This is not an unusual or unknown situation in environmental pollution management.

For our present case study, the choice of both the indicator and the red-line is straightforward:

**Case study (2/6):** The Starlight Foundation requires, for its Tourism Destination certification, fulfilling a series of requirements regarding night sky brightness, atmospheric transparency, and seeing. Night sky brightness can be monitored in the SQM band, and the threshold value (red-line) is set at 21.00 $mag_{SQM}/arcsec^2$.

*4.3. Evaluating present compliance (or lack thereof)*

Once the red-lines have been set, the next step is evaluating the actual situation. This requires monitoring the relevant variables and assessing how far from the limits (above or below) we are.

**Case study (3/6):** The light pollution levels for our case study can be determined from the zenithal night sky brightness records of the MeteoGalicia SQM detector located in the Cíes islands, latitude 42.2118º, longitude –8.9084º (WGS84, EPSG:4326), and an altitude of 25 m above sea level. This is one of the 25 stations where SQM are installed, belonging to the global monitoring network of MeteoGalicia, the Galician public meteorological agency (Bará, 2016; MeteoGalicia, 2020). They record the night sky brightness at a rate of one sample per minute. The night sky brightness varies due to multiple factors, with characteristic timescales from seconds to years. The typical sky brightness in clear and moonless nights is well described by the so-called $m_{FWHM}$ magnitude (Bará et al, 2019b), defined as the average value of the SQM records contained within the full width at half-maximum interval of the clear nights' aerosol-driven peak, under "no-Sun no-Moon conditions" (Sun below −18° and Moon below −5°).

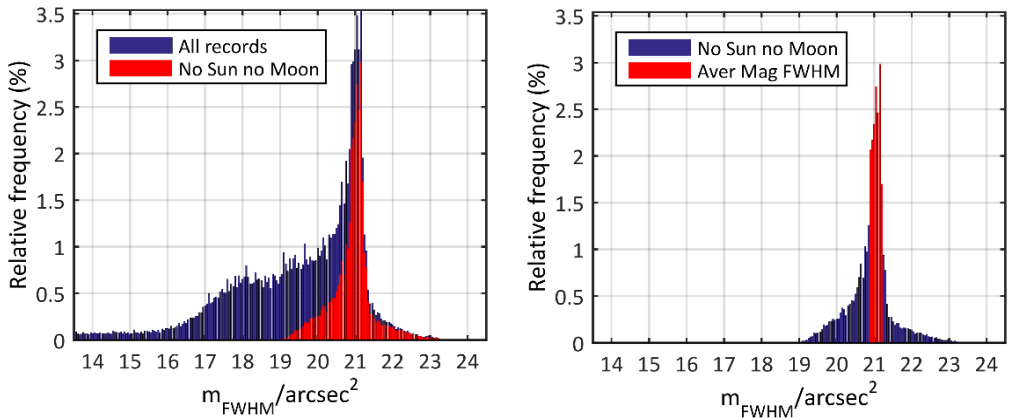

**Fig. 2**. Histograms of zenithal night sky brightness in the SQM band corresponding to the year 2018 in the MeteoGalicia weather station of Cíes islands. Left: all values larger than 13.5 $mag_{SQM}/arcsec^2$, and values under 'no-Sun, no-Moon' conditions (Sun below −18° and Moon below −5°); Right: 'No-Sun, no-Moon' histogram, and histogram of the values contained within the full-width at half maximum region of the clear nights' peak. The $m_{FWHM}$ is the average of the values contained in this region of the histogram.

The region of the brightness histogram where the m$_{FWHM}$ is calculated is shown in red in the right panel of Fig. 2. This metric excludes the effects of the small wing of extremely high darkness records due to dense fog episodes in this coastal area. The nominal m$_{FWHM}$ recorded in this station in the year 2018 was 21.06±0.03 mag$_{SQM}$/arcsec$^2$ (one-sigma combined uncertainty). This means that the zenith night sky over this island (evaluated with this metric) was 0.06 mag$_{SQM}$/arcsec$^2$ darker than the limiting red-line (21.00), that is, there was a 6% margin of allowable increase in the night sky brightness before reaching the critical value that would put the Tourism Destination certification of this National Park at risk.

*4.4. Evaluating preventive or remediation options*

After checking compliance (or lack thereof) with the red-lines, the margin before attaining the limit of allowable emissions (or the size of the required reductions, in case of failure) shall be critically analyzed and put into context. This provides a first insight about the dimension of the challenge to be addressed.

> **Case study (4/6):** It shall be kept in mind that the nominal brightness of the night sky over the island in 2018, m$_{FWHM}$=21.06±0.03 has contributions from both light pollution and natural light from the sky (Masana et al, 2021). Assuming for the purposes of this exercise a reference natural sky brightness of 22.00 mag$_{SQM}$/arcsec$^2$ we have that the artificial sky radiance over the island in the SQM band, given by the difference between the total and the natural radiances, was $L_a = 10^{-0.4 \times 21.06} - 10^{-0.4 \times 22.00} = 21.82 \times 10^{-10}$, in arbitrary linear units. In the same units scale, the maximum allowable sky radiance compliant with the red-line of 21.00 mag$_{SQM}$/arcsec$^2$ would be $L_{a,max} = 10^{-0.4 \times 21.00} - 10^{-0.4 \times 22.00} = 23.96 \times 10^{-10}$. This implies that the maximum allowable artificial radiance is $L_{a,max}/L_a = 1.098$ times the actual one. In other words, if the artificial sky brightness over the Cíes islands would increase by a ~10% over its present nominal value, the red-lines will be clearly surpassed.

Preventing the red-lines from being crossed implies, for our case study, including in the mid-term planning of the territory the strict requirement that the absolute, distance-weighted emitted radiant flux cannot increase by more than a 10% on average in the foreseeable future, including the period of time for which protection against light pollution shall be granted to the National Park. This overall average can be distributed territorially in many alternative ways, taking into account that each pixel emits a different amount of radiance and that this radiance contributes more or less efficaciously to the zenith sky brightness depending on several factors, as e.g. the distance to the observation point, as analyzed in the section 4.5.

*4.5. Allocating quotas and duties among intervening actors*

Once the margins for additional emissions (or the required reductions in case red-lines have already been surpassed) are determined, they should be translated into operative limits for all social agents responsible for decisions on outdoor lighting. These social agents

(municipalities, owners of privately lit premises, etc.) must be clearly identified, and their relative contribution to the present light pollution levels shall be established. The calculation of by how much each one contributes to the artificial brightness should be made with the accuracy and precision required for enabling the adoption of science-informed public decisions on lighting. Note that these accuracy and precision are generally much lower than the ones required to test scientific theories and models of light pollution propagation. What matters is to determine where the main light pollution sources are and to have a reasonable estimate of their percent contribution to the overall light pollution values at the observing site.

The calculation of the contribution of every pixel of the surrounding territory can be made using available models (Aubé et al, 2020; Bará and Lima, 2018; Falchi and Bará, 2020). This usually requires standardizing the atmospheric conditions under which will be carried out the assessment. As in other choices of this kind, the particular standard conditions shall be consistent with the prevailing ones, and the light propagation models shall ideally include all relevant physical processes that are into play.

The physics of the light propagation in the atmosphere at the usual radiance levels of outdoor lighting is for all practical purposes linear. This means that every pixel of the territory contributes to the artificial sky brightness in proportion to its absolute radiance emissions, being the constant of proportionality dependent on the angular emission pattern of the light sources, their spectral radiant density, the spectral reflectance of the pavements, the presence of obstacles blocking the propagation of light along some set of rays, the composition and concentration profiles of the molecular and aerosol atmosphere, and, of course, on the distance from the emitting sources to the observation point. A wide set of models for characterizing this propagation are available in the literature (Aubé, 2015; Aubé and Simoneau, 2018 Aubé et al, 2020; Cinzano and Falchi, 2012; Garstang, 1986, 1989, 1991; Kocifaj, 2007, 2016, 2018; Kocifaj and Bará, 2019). They basically provide the light pollution point spread function (PSF), that is the contribution of a unit radiance source to the sky radiance as a function of the distance and wavelength (with the remaining variables mentioned above acting as parameters of the model). Once the two-dimensional PSF and the ground distribution of sources is known, the contributions of the surrounding territory to the brightness at the observation point can be added up in suitable administrative or functional areas and displayed for analysis, as in Figure 3.

> **Case study (5/6):** For the purposes of this exercise, we used the PSF for the zenithal sky brightness in the V band calculated by Cinzano and Falchi (2012), assuming a layered atmosphere with a clarity parameter K=1 (visibility 26 km). We calculated the contributions of each individual pixel to the zenith brightness at the observation point on the island, as well as the contributions aggregated by municipalities, the main administrative and political bodies responsible for public lighting in Galicia. The results are shown in Fig. 3.

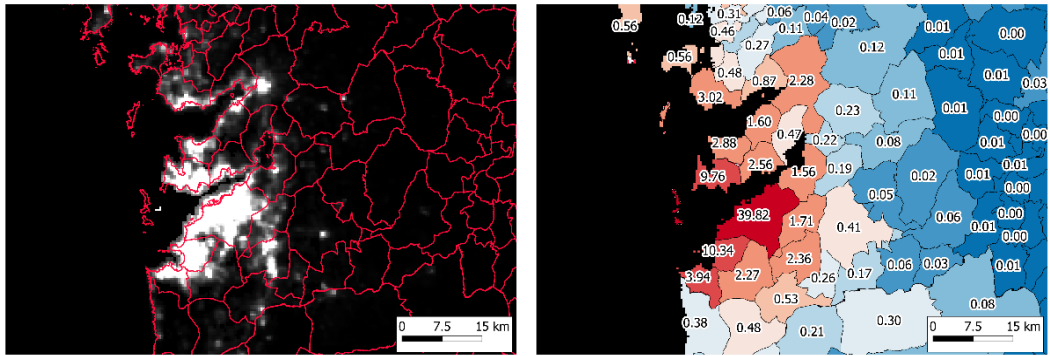

**Fig. 3.** Left: Artificial weighted sources map, obtained as the pixel-wise product of the two-dimensional PSF by the VIIRS-DNB raw sources (Figure 1, right). The gray level of each pixel is proportional to its contribution to the zenithal sky radiance over the Cíes island SQM detector. The bright pixels in the sea at the West of the islands correspond to a waiting anchorage area for large carrier ships bound to Vigo harbor; Right: zenithal sky brightness contributions aggregated by municipalities. The main contributor to the island light pollution is the municipality of Vigo (39.82%), followed by Nigrán (10.34%), Cangas (9.76%), and many other municipalities with lower percentages.

The municipalities' contributions map shown in Fig. 3 (right) shows a typical situation for protected sites located near highly populated urban areas. A few municipalities, the most populated and close ones, contribute with the larger percent share to the deterioration, followed by several dozen ones contributing with progressively smaller amounts. Note that in typical rural regions with no large cities nearby, the individual contributions of the municipalities tend to be smaller, and the number of relevant contributors tend to be larger, extending to longer distances (Bará and Lima, 2018).

The map in Fig. 3 also reveals the multiple-choice scenario compatible with the prevention goals. If, hypothetically, authorizing new emissions would be a desirable option (which, in the opinion of the authors, is strongly discouraged from a conservation standpoint), one could trivially grant to every municipality an additional 10% emission quota above their present emission values. But this uniform rate is probably not the optimum choice when social needs, priorities and inter-territorial solidarity are included in the mix. Some municipalities could be granted larger additional emissions, at the expense of others being granted less, consistent with the weightings displayed in Fig. 3 and ensuring that the effect of these emissions do not attain, in any case, a 10% increase in the final indicator (the artificial brightness over the islands, addressed in Case study subsections 4 to 6).

*4.6. Steer the transition process*

Realizing in practice this approach requires an important effort of mid-term territorial planning, encompassing multiple social and environmental factors. Transparent, participated, science-informed, and fully democratic decisions may provide the necessary support for a decided public action to preserve or remediate the situations of concern. Permanent monitoring of the relevant variables, e.g. including the night sky brightness as a relevant environmental parameter routinely monitored by meteorological agencies (Bará, 2016; Bertolo et al, 2019; MeteoGalicia, 2020) is an essential tool. Follow-up reports with updated data and evaluation of potential threats or opportunities are an indispensable tool in support of public decision making.

*4.7. What if the red-lines were already surpassed?*

In many places of the world the light pollution levels have already surpassed any reasonable red-line. It is required, then, to elaborate, approve, and carry out a suitable remediation process to reduce the light pollution levels until they fall below the maximum admissible values. The overall approach to follow in this important case is essentially the same as the one described in the subsections above for avoiding surpassing the limits, but with reversed signs. In the remediation case, the required target indicator values have to be attained by reducing the emissions in the surrounding territory. Percent contribution maps as the one displayed in the right side of Figure 3 are instrumental tools to assess by how much a given reduction in one municipality will contribute to reduce the overall light pollution at the protected site. If red-lines were surpassed, absolute reductions in emissions are strictly unavoidable. All conditions mentioned in section 4.6 for steering the transition process apply specially in this case. This reduction process may arise also when there is a revision of the red-lines, due to the will to better protect a place against the detrimental effects of light pollution. In our example, the Galician Atlantic Islands Maritime-Terrestrial National Park may decide that the islands need a higher level of protection, such as that to fulfill the more stringent requirements for the Starlight Reserve certification.

## 5. Additional remarks

We described in this work a simplified but workable approach for enabling light pollution management by means of outdoor lighting territorial planning. The case study presented here is intended as an example of application to show how could this approach be put in practice. It should not be taken as a definite prescription for other locations or environmental problems. For instance, in many cases the artificial zenith night sky brightness in the SQM band will not be the most adequate indicator. Other artificial radiance indicators should be used for assessing global nightscape or ecological effects, both from the geometrical viewpoint (e.g. average radiance of the upper hemisphere, average radiance below 10° above horizon, horizontal irradiance, azimuthally-averaged vertical irradiance...) and from the spectral one (using ecological spectral sensitivity bands different from the SQM). The overall way of approaching the problem and the main steps to be carried out do not change, though.

## 6. Conclusion

A light pollution control approach exclusively based on establishing conditions upon individual light sources (ULOR, spectrum, illuminance levels, glare, ...), seems to be clearly insufficient by itself for curbing the actual process of degradation of the night. A complementary, not substitutive, 'red-lines' strategy should in our opinion be adopted as early as possible by the light pollution community. This top-down approach is based on agreeing definite limits on the maximum allowable degradation of the night, operationally given as quantitative indicator limits. Numerical models can be used to determine the contribution of each patch of the surrounding territory, and the lighting system installed therein, to the light pollution levels at the sites of interest. In combination with a clearly defined set of red-lines, the percent contribution maps provide a key methodological tool for science-informed public policy decision-making. Light pollution abatement should routinely be included as an integral part of territorial planning by all concerned administrative bodies.

**Acknowledgments**

This work was supported by Xunta de Galicia, grant ED431B 2020/29. CITEUC is funded by National Funds through FCT - Foundation for Science and Technology (project: UID/MULTI/00611/2019) and FEDER - European Regional Development Fund through COMPETE 2020 – Operational Programme Competitiveness and Internationalization (project: POCI-01-0145-FEDER-006922).

**CRediT author statement**

SB: Conceptualization, Methodology, Software, Data curation, Draft preparation; FF: Conceptualization, Methodology, Software, Writing-Reviewing and Editing; RCL: Conceptualization, Writing-Reviewing and Editing; MP: Conceptualization, Writing-Reviewing and Editing.

**Conflicts of interest**

The authors declare no conflicts of interest.